\documentclass[a4paper,12pt]{article}

\usepackage{indentfirst}
\usepackage{hyperref}
\usepackage{multirow}
\usepackage{latexsym,bm,amsmath,amssymb}
\usepackage{graphicx}
\usepackage{epsfig}
\usepackage{subfigure}
\usepackage{cite}
\usepackage{color}
\usepackage[top=1in, bottom=1in, left=1.2in, right=1.2in]{geometry}
\usepackage{appendix}
\usepackage{ulem}
\usepackage{slashed}

\begin{document}

\begin{center}
{\large\bf Unitarity Analyses of $\pi N$ Elastic Scattering Amplitudes}

{Y.F. Wang$^{1,\P}$}

$^1${Department of Physics and State Key Laboratory of Nuclear Physics and Technology,\\Peking University, No. 209, Chengfu Road, Haidian district, Beijing 100871, Beijing, PR China}

$^\P${E-mail: 1401110076@pku.edu.cn}
\end{center}

\centerline{\bf Abstract}
The pion - nucleon scattering phase shifts in $s$ and $p$ waves are analyzed using PKU unitarization approach that can separate the phase shifts into different contributions from poles and branch cuts. It is found that in $S_{11}$ and $P_{11}$ channels, there exist large and positive missing contributions when one compares the phase shift from known resonances plus branch cuts with the experimental data, which indicates that those two channels may contain sizable effects from $N^*(1535)$ and $N^*(1440)$ shadow poles. Those results are obtained using tree level results of the $\pi N$ amplitude. 

The $\pi N$ elastic scattering is one of the most fundamental and important processes in nuclear or hadron physics. The research on this subject lasts for decades till today, but there are still some opening questions. For recent studies, one refers to for example, Refs.\cite{Becher:2001hv,Ditsche:2012fv,Alarcon:2011zs,Chen:2012nx}. Also, the intermediate resonances such as $N^*(1535)$ and $N^*(1440)$ have attracted sustained attention. For $N^*(1535)$, the origin of its high mass and its large coupling to the $\eta N$ channel need to be studied further\cite{Kaiser:1995cy,Nieves:2001wt}. As for $N^*(1440)$, its quark model interpretation and its coupling to $\sigma N$ channel are still not well understood\cite{Krehl:1999km}; besides, $N^*(1440)$ may contain a two-pole structure\cite{Arndt:1985vj}, and the corresponding $P_{11}$ channel may have strange branch cuts in the complex $s$ plane\cite{Ceci:2011ae}. Therefore, a method is needed to exam the relevant channels carefully and to exhume more physics behind.

In this talk we use a new unitarization method, called PKU representation\cite{Zheng:2004kappa,Zhou:2006wm,Xiao:2001pku,Zhou:2004ms}, to study $\pi N$ elastic scatterings. This method provides a production representation of the partial wave $S$ matrix for a $2\to2$ elastic scattering process (the masses of the two particles are $m_1$ and $m_2$ respectively):
\begin{equation}\label{PKU}
S(s)=\prod_b \frac{1-\text{i}\rho(s)\frac{s}{s-s_l} \sqrt{\frac{s_b-s_l}{s_r-s_b}}}
{1+\text{i}\rho(s)\frac{s}{s-s_l} \sqrt{\frac{s_b-s_l}{s_r-s_b}}}
\prod_v \frac{1+\text{i}\rho(s)\frac{s}{s-s_l} \sqrt{\frac{s_v'-s_l}{s_r-s_v'}}}
{1-\text{i}\rho(s)\frac{s}{s-s_l} \sqrt{\frac{s_v'-s_l}{s_r-s_v'}}}
\prod_r \frac{M^2_r-s+\text{i}\rho(s)sG_r}{M^2_r-s-\text{i}\rho(s)sG_r}
e^{2\text{i}\rho(s)f(s)}\ \mbox{. }
\end{equation}
In Eq.(\ref{PKU}), the kinematic factor $\rho(s)=\frac{\sqrt{s-s_l}\sqrt{s-s_r}}{s}$, with $s_l=(m_1-m_2)^2$, $s_r=(m_1+m_2)^2$; the functions corresponding to the resonance terms are
\begin{align}
&M^2_r=\text{Re}[z_r]+\text{Im}[z_r]\frac{\text{Im}[\sqrt{(z_r-s_r)(z_r-s_l)}]}{\text{Re}[\sqrt{(z_r-s_r)(z_r-s_l)}]}\ \mbox{, }\\
&G_r=\frac{\text{Im}[z_r]}{\text{Re}[\sqrt{(z_r-s_r)(z_r-s_l)}]}\ \mbox{, }
\end{align}
and $s_b,\ s_v'$ and $z_r$ denote bound state poles (on the first Riemann sheet real axis below threshold), virtual state poles (on the second Riemann sheet real axis below threshold) and resonances (on the second Riemann sheet off the real axis), respectively. Lastly, the exponential term in Eq.(\ref{PKU}) is named as ``background term'' since it contains no poles. Actually, the background term carries the information of left-hand cuts and even inelastic cut above inelastic thresholds, and it satisfies a dispersion formula:
\begin{equation}\label{fdisper}
f(s)=\frac{s}{2\pi\text{i}}\int_{lhc.}ds'\frac{\text{disc}f(s')}{(s'-s)s'}+\frac{s}{2\pi\text{i}}\int_{inel.}ds'\frac{\text{disc}f(s')}{(s'-s)s'} \ \mbox{, }
\end{equation}
where the ``$lhc.$'' and ``$inel.$'' denote the left-hand cut and inelastic cut respectively, and ``disc'' stands for the discontinuity of the function $f(s)$ on the cuts. In fact, Eq.(\ref{fdisper}) is under a one-order substraction with the substraction point $s_0=0$. Due to a general fact that $f(0)=0$\cite{Zhou:2006wm}, the substraction constant term $f(s_0)$ vanishes in Eq.(\ref{fdisper}).

The PKU representation provides additive phase shift from different contributions, which makes the analysis of phase shift clear and convenient. Moreover, each phase shift contribution has definite sign: bound states always give negative contributions, while virtual states and resonances always give positive contributions; the left hand cuts would empirically give negative phase shifts. Furthermore, the phase shifts given by PKU representation are sensitive to (not too) distant poles. Because of the advantages above, this method has successfully been used in $\pi\pi$ and $\pi K$ elastic scatterings\cite{Xiao:2001pku,Zheng:2004kappa,Zhou:2004ms} and to check the reliability of some traditional unitarization approaches\cite{Zheng:2004kappa}.

In this talk, PKU representation is used on $s$ and $p$ wave channels of the $\pi N$ elastic scattering, and to check traditional $K$-matrix unitarization method about the spurious pole issue. When one uses PKU representation as a unitarization method, the left-hand cut contribution can be calculated via perturbation theory, while the pole positions are added as input parameters. Here we calculate the process using baryon chiral perturbation theory only at tree level, and omit the inelastic cut since the energy region to be analysed is far from the inelastic thresholds. The $\mathcal{O}(p^1)$ and $\mathcal{O}(p^2)$ $\pi N$ Lagrangians in $SU(2)$ case can be found in Ref.\cite{Fettes:2000gb}, while we use the same isospin and helicity decompositions as Ref.\cite{Chen:2012nx}.

At tree level $\pi N$ elastic scattering the left-hand cut structure is quite simple: only one kinematic cut $(-\infty,0]$ and one cut $[(M^2-m^2)^2/M^2,2m^2+M^2]$ due to the $u$ channel nucleon exchange. Actually the contribution from the latter is numerically very small, so the dominant contribution is
\begin{equation}
f(s)=\frac{s}{\pi}\int_{-\infty}^0 \frac{\sigma(w)dw}{w(w-s)}\ \mbox{, }
\sigma(w)=\text{Im}\Big\{\frac{\ln|S_{\text{tree}}|}{2 i\rho(w)}\Big\}\ \mbox{, }
\end{equation}
which is always negative, since $\rho(w)>0$ when $w<0$ and $|S_{\text{tree}}|>1$ in perturbation theory.

As for poles, however, the PKU method can only deal with the two-sheet structure of elastic scatterings, so what we need are resonances on the second sheet (called shadow poles) rather than the third sheet poles given by experiments (i.e. $\sqrt{s}^{\text{III}}=M_{\text{pole}}-\frac{i}{2}(\Gamma_{\text{inelastic}}+\Gamma_{\pi N})$). Generally second sheet shadow poles have no direct relation to the third sheet resonances, but under narrow width approximation, the shadow pole positions (if exist) may be estimated as
\begin{equation}
\sqrt{s}^{\text{II}}=M_{\text{pole}}-\frac{i}{2}(\Gamma_{\text{inelastic}}-\Gamma_{\pi N})\ \mbox{. }
\end{equation}
In the calculation the resonance parameters are taken from Ref.\cite{Anisovich:2011fc}.

\begin{figure}[]
\center
\subfigure[]{
\label{com:subfig:S11}
\scalebox{1.0}[1.0]{\includegraphics[width=0.3\textwidth]{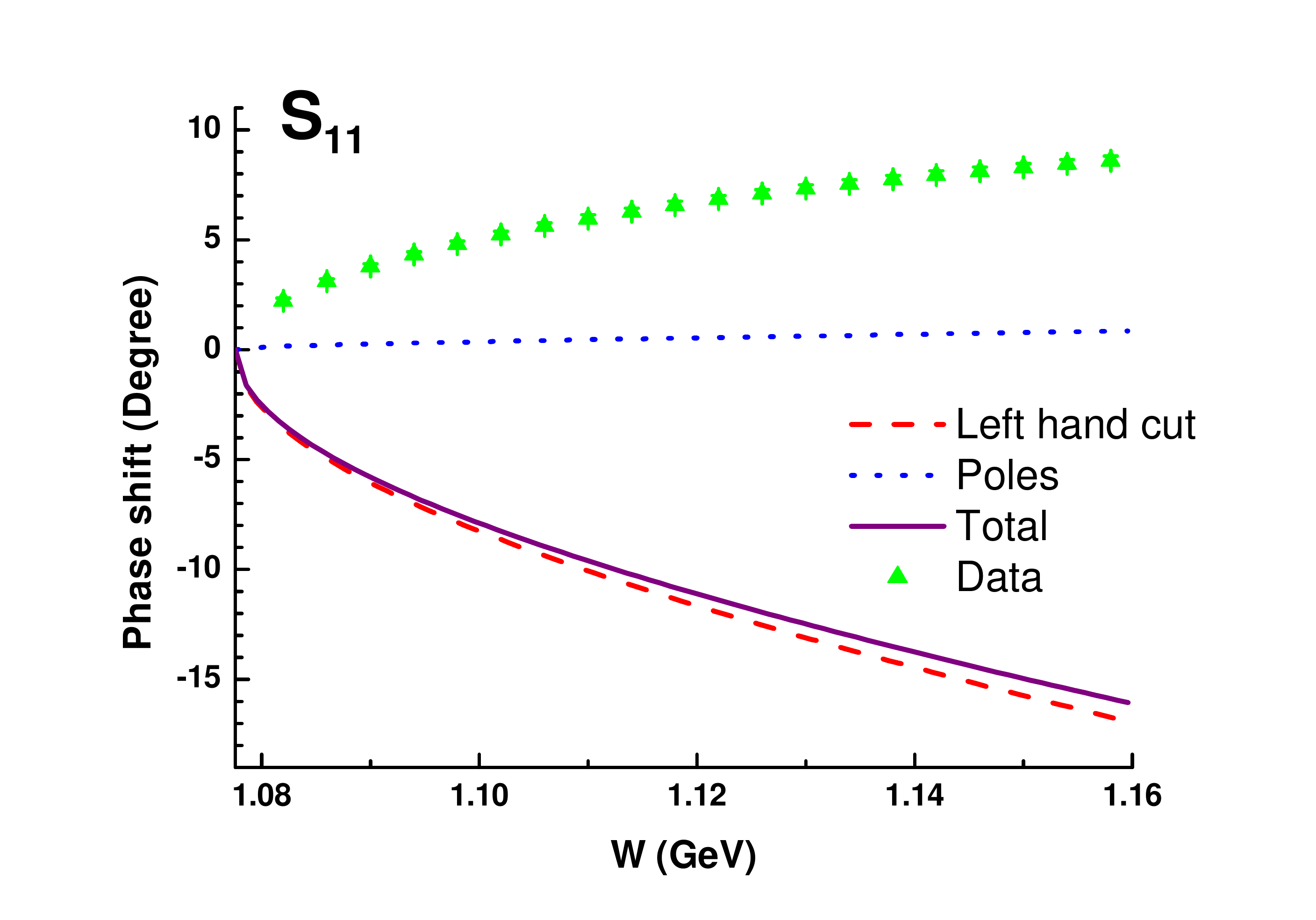}}}
\subfigure[]{
\label{com:subfig:S31}
\scalebox{1.0}[1.0]{\includegraphics[width=0.3\textwidth]{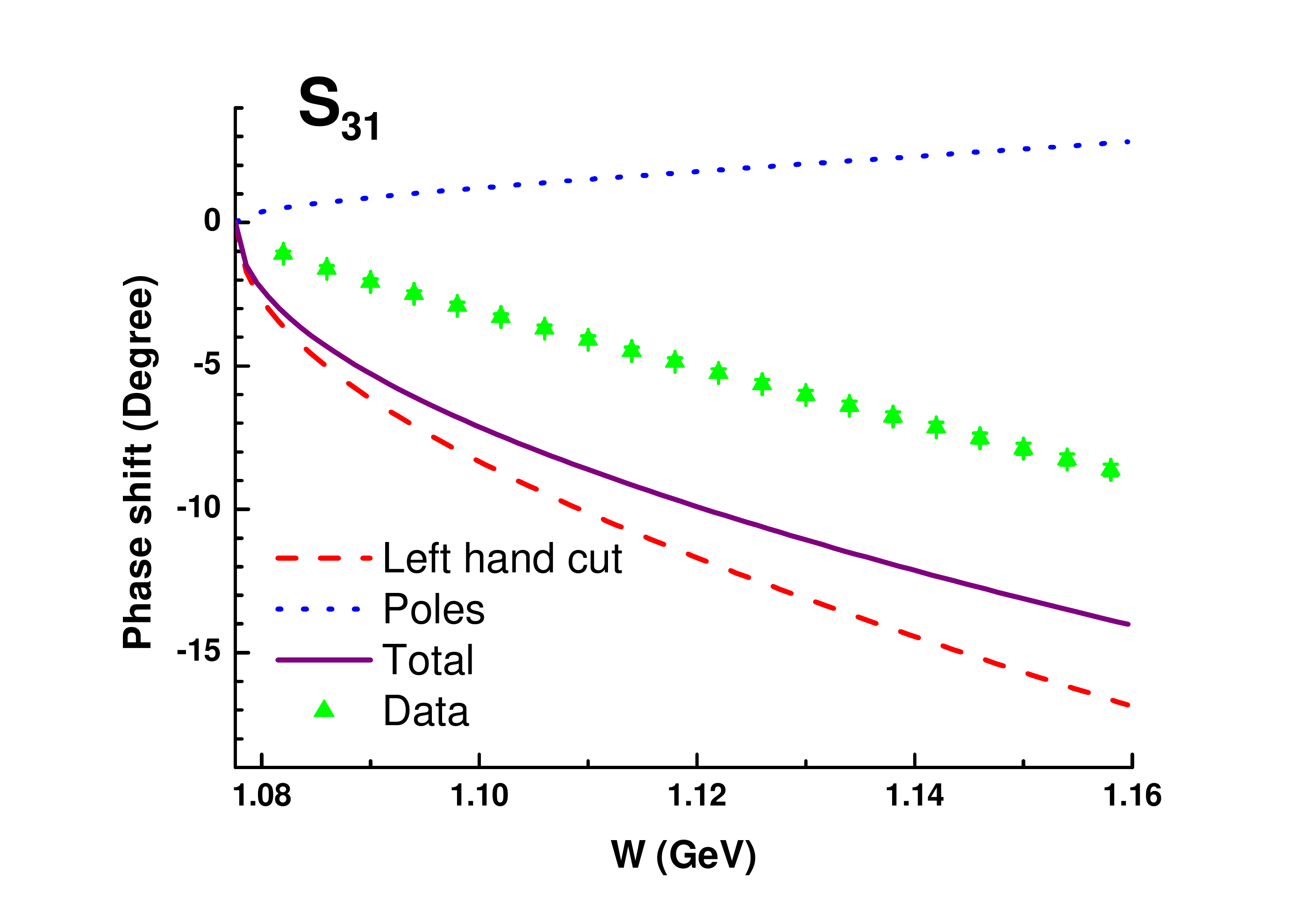}}}
\subfigure[]{
\label{com:subfig:P11}
\scalebox{1.0}[1.0]{\includegraphics[width=0.3\textwidth]{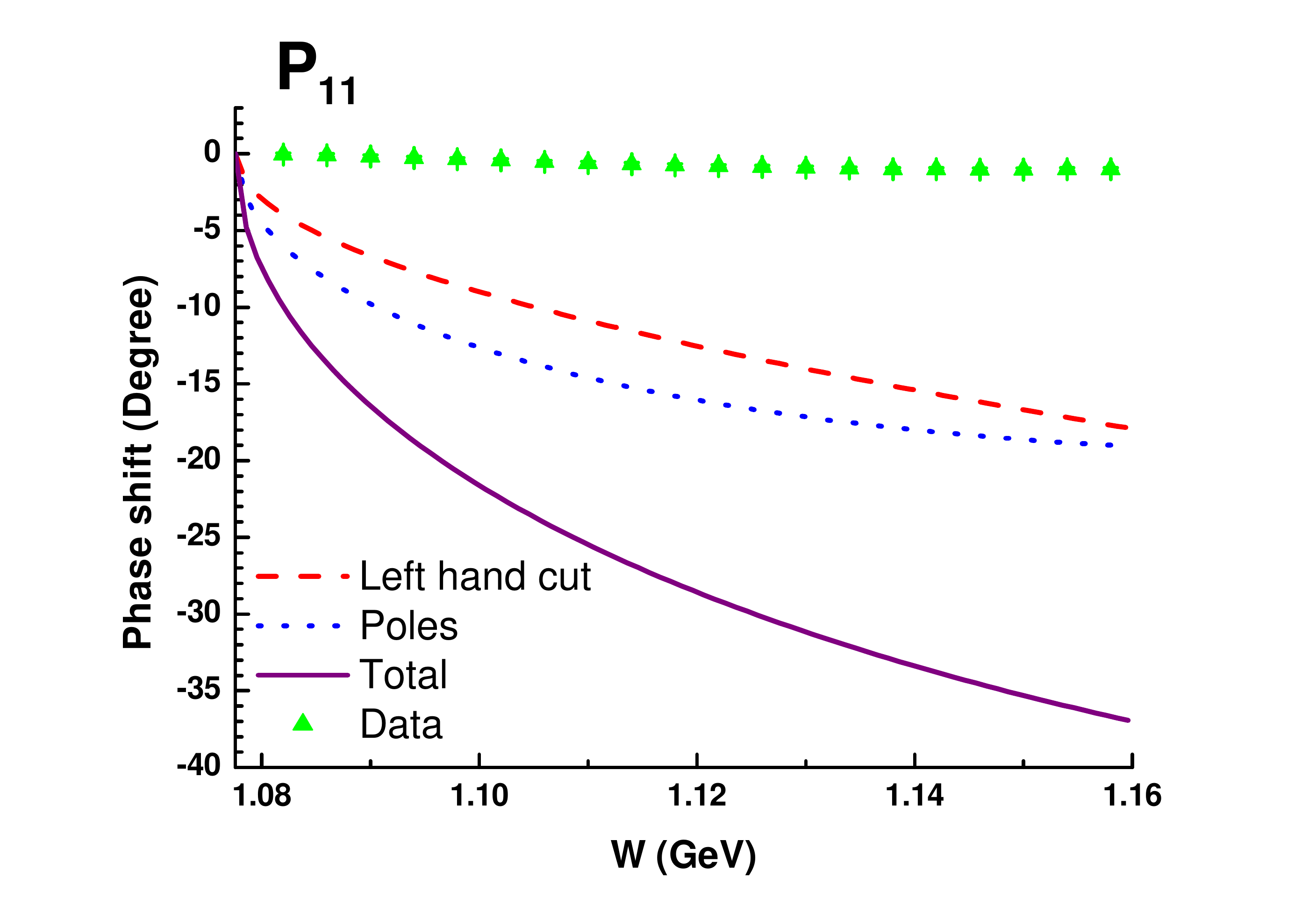}}}
\subfigure[]{
\label{com:subfig:P31}
\scalebox{1.0}[1.0]{\includegraphics[width=0.3\textwidth]{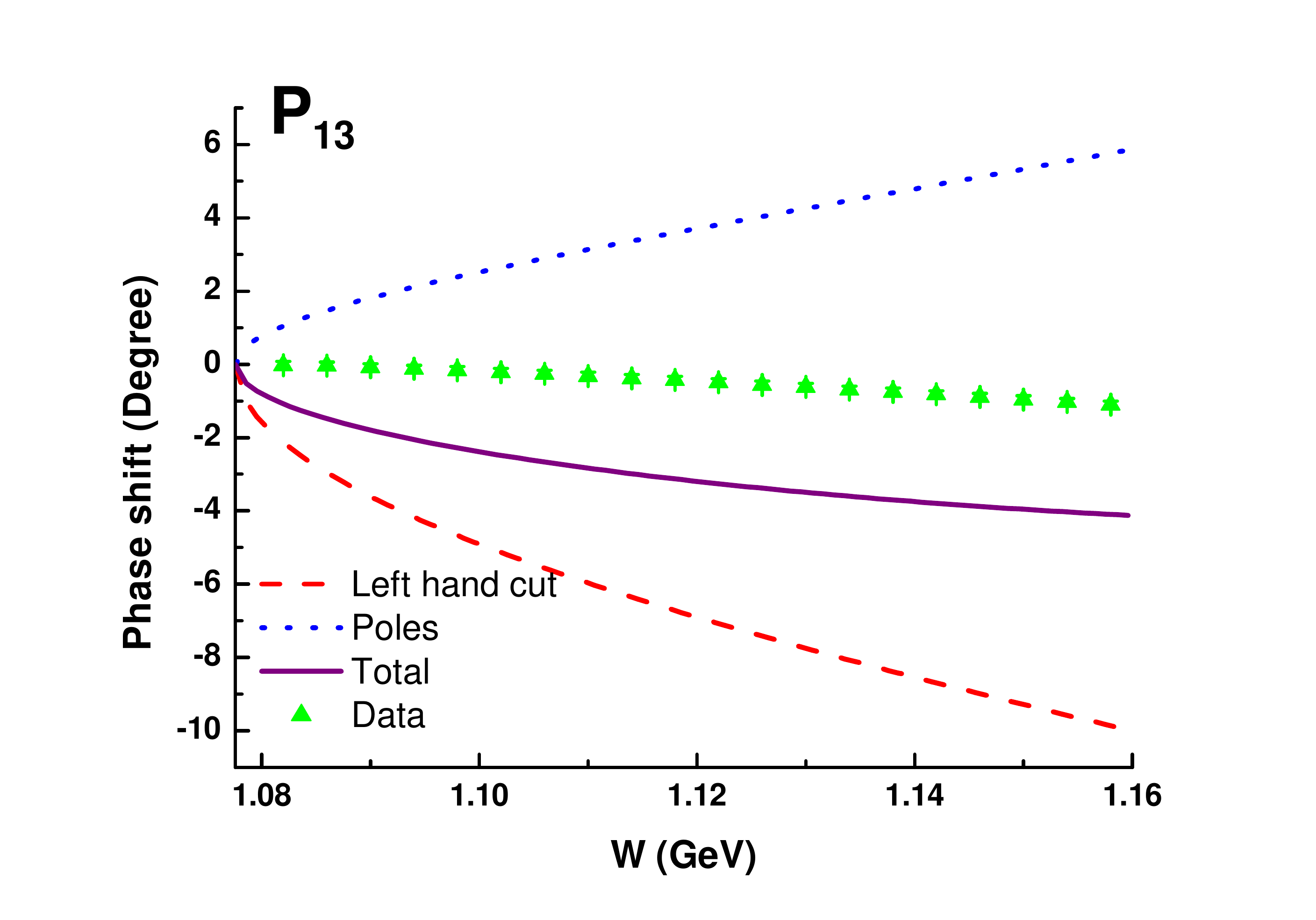}}}
\subfigure[]{
\label{com:subfig:P13}
\scalebox{1.0}[1.0]{\includegraphics[width=0.3\textwidth]{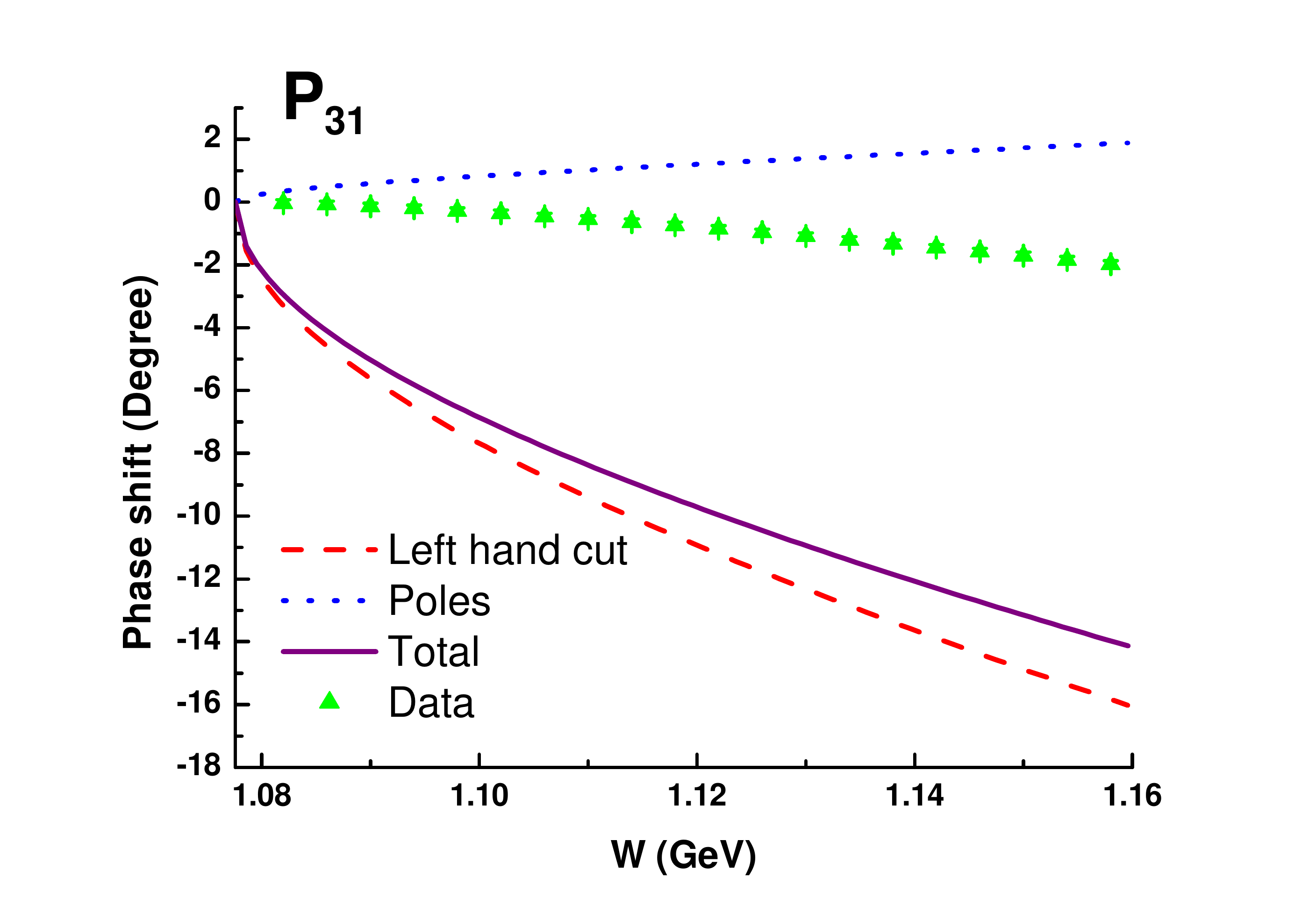}}}
\subfigure[]{
\label{com:subfig:P33}
\scalebox{1.0}[1.0]{\includegraphics[width=0.3\textwidth]{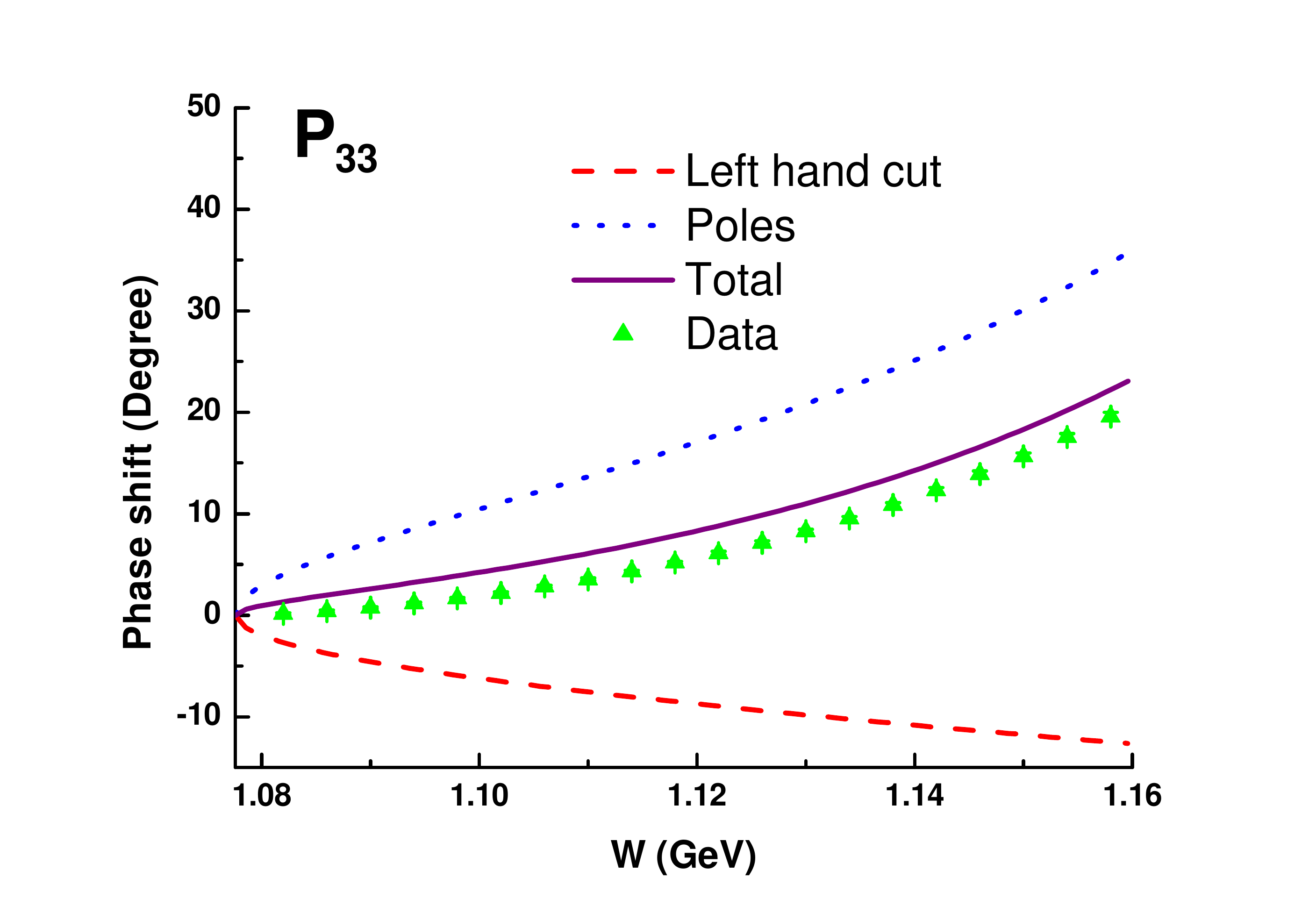}}}
\caption{Tree level PKU representation analyses of the $\pi N$ elastic scattering in $s$ and $p$ waves. }\label{p2PKU}
\end{figure}

The masses and $\mathcal{O}(p^1)$ coupling constants are set as $M=0.9383\text{GeV}$, $m=0.1396\text{GeV}$, $F=0.0924\text{GeV}$ and $g=0.1267$, and the $\mathcal{O}(p^2)$ coupling constants are determined by a rough $K$-matrix fit to the data in Ref.\cite{Arndt:2006data} and estimated errors in Ref.\cite{Chen:2012nx}. The tree level $K$-matrix formula is
\begin{equation}\label{Kfit}
T_K=\frac{T_{\text{tree}}}{1-i\rho T_{\text{tree}}}\ \mbox{, }\delta_K=\arctan\big[\rho T_{\text{tree}}\big]\ \mbox{. }
\end{equation}
We fit $20$ data in each channel, getting the results $c_1=-0.642,\,c_2=1.559,\,c_3=-2.812,\,c_4=1.560$, with the fit quality $\chi^2/\text{d.o.f}=3.624$. Then the different contributions to phase shifts in each channel are plotted in Fig.\ref{p2PKU}. It's found that especially in $S_{11}$ and $P_{11}$ channels, just the two where $N^*(1535)$ and $N^*(1440)$ stay, there exists a huge disagreement between the phase shifts from ``known'' poles plus cut and the data: the former contributions have missed some important positive contributions. There are some possible interpretations of that: firstly it is natural to think that the one loop contributions in the two channels may be notable; secondly, the previous calculation of the shadow poles is only under a rough approximation which might give wrong positions; thirdly, we can not exclude the existence of some hidden poles, i.e. virtual states or resonances below the threshold; finally, other branch cuts which are not included in the discussions above may be non-ignorable, e.g. the inelastic cut and the type of cut proposed by Ref.\cite{Ceci:2011ae}. All in all, that disagreement may indicate some special dynamical properties of $N^*(1535)$ and $N^*(1440)$.

We have also tried to find the shadow pole positions under the assumption that hidden poles or extra cuts do not attend in the $S$ matrix, getting a bewildering result that the shadow pole positions may lie below $\pi N$ threshold: $M_{1535}=0.880$GeV and the width $\Gamma_{1535}=540$MeV, while $M_{1440}=0.896$GeV and $\Gamma_{1440}=14.6$MeV. The fit qualities are $\chi^2/\text{d.o.f}=0.017,\ 3.738$, respectivly. Notice that the fit for $P_{11}$ channel obeys a constraint that $\mathcal{O}(|\mathbf{k}|^1)$ term in the phase shift vanishes.

\begin{figure}[]
\center
\includegraphics[width=0.4\textwidth]{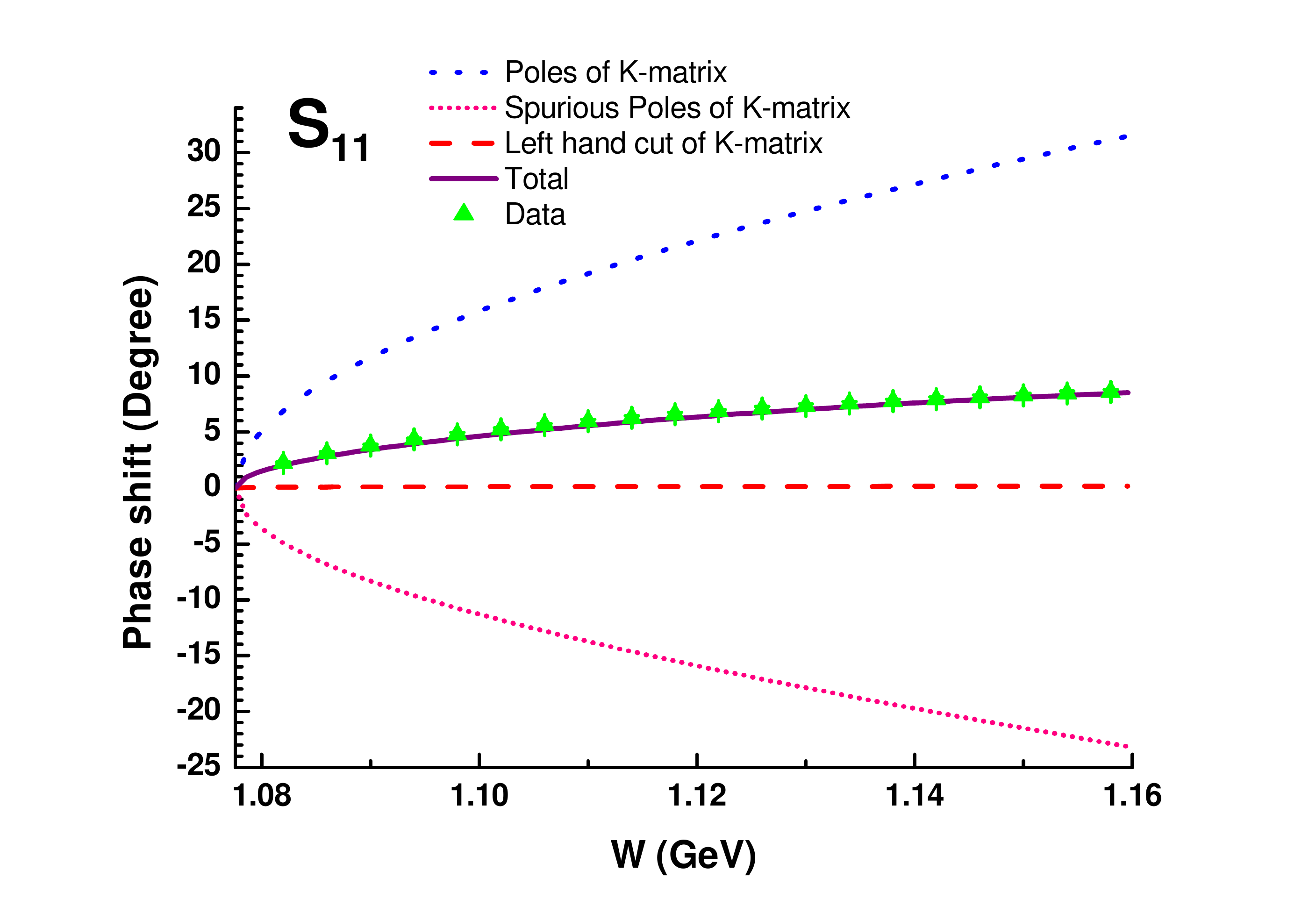}
\caption{The $S_{11}$ channel $K$-matrix fit in PKU representation. }
\label{fig:S11p2K}
\end{figure}

Finally, inspired by Ref.\cite{Zheng:2004kappa}, we applied the PKU representation method to the $K$-matrix amplitude Eq.(\ref{Kfit}), to examine the quality of the traditional $K$-matrix fit. Unfortunately, we have indeed found that the $K$-matrix fit gives some spurious poles on the first sheet off the real axis. And worse yet, each phase shift given by $K$-matrix suffers from a cancellation between a large negative contribution from spurious poles and one from second sheet poles, leaving ridiculously tiny left-hand cut contributions. For example, the $S_{11}$ channel result is shown in Fig.\ref{fig:S11p2K}. Hence $K$-matrix method is not reliable in dynamically generating poles at least at tree level, despite the fact that it is widely used to determine the coupling constants.

To summarize, we apply the PKU representation which are derived from first principles, to analyse the $\pi N$ elastic scattering processes. It is found that in $S_{11}$ and $P_{11}$ channels there exist some large missing positive contributions if we only consider the known poles and perturbative left-hand cuts. Our preliminary results are very encouraging, since it uncovers some hidden truths related to $N^*(1535)$ and $N^*(1440)$. In the future, we should examine different possibilities of the missing positive contributions, and do more careful quantitative analyses.
\section*{Acknowledgment}
I would like to thank Han-Qing Zheng and De-Liang Yao for helpful advices. This work is supported in part by National Nature Science Foundations of China (NSFC) under Contracts No. 10925522, No. 11021092, No. 11575052 and No. 11105038.

\end{document}